\UseRawInputEncoding
\documentclass[twocolumn,prc,superscriptaddress,unsortedaddress,preprintnumbers,amsmath,amssymb]{revtex4}
\usepackage{tipa}
\usepackage{float}
\usepackage{color}
\usepackage[colorlinks,
			linkcolor=blue,
			anchorcolor=blue,
			citecolor=blue]{hyperref}
\usepackage{amsmath}
\definecolor{gray}{RGB}{220,220,220}

\usepackage{graphicx}
\usepackage{dcolumn}
\usepackage{bm}

\def\beq{\begin{equation}}
	\def\eeq{\end{equation}}
\def\bea{\begin{eqnarray}}
	\def\eea{\end{eqnarray}}

\def\fun#1#2{\lower3.6pt\vbox{\baselineskip0pt\lineskip.9pt
		\ialign{$\mathsurround=0pt#1\hfil##\hfil$\crcr#2\crcr$\sim$\crcr}}}
\preprint{}
\begin{document}
    \title{Phase diagrams of neutron-proton superfluid in asymmetric nuclear matter}
	\author{K. D. Duan}
        \affiliation{Department of Physics, Hebei University, Baoding, 071002, China}
	\affiliation{CAS Key Laboratory of High Precision Nuclear Spectroscopy, Institute of Modern Physics, Chinese Academy of Sciences, Lanzhou 730000, China}
    \affiliation{School of Nuclear Science and Technology, University of Chinese Academy of Sciences, Beijing 100049, China}
        \author{H. B. Zhang}
        \affiliation{Department of Physics, Hebei University, Baoding, 071002, China}
        \affiliation{Key Laboratory of High-Precision Computation and Application of Quantum Field Theory of Hebei Province, Baoding, 071002, China}
        \author{X. L. Shang}\email[]{shangxinle@impcas.ac.cn}
	\affiliation{CAS Key Laboratory of High Precision Nuclear Spectroscopy, Institute of Modern Physics, Chinese Academy of Sciences, Lanzhou 730000, China}
    \affiliation{School of Nuclear Science and Technology, University of Chinese Academy of Sciences, Beijing 100049, China}
	\begin{abstract}
        The finite-temperature phase structures for neutron-proton superfluidity in asymmetric nuclear matter are investigated, with a particular focus on the angular dependence of the pairing gap induced by the $^3SD_1$ $NN$ interaction. This angular dependence of the pairing gap results in the Cooper pair momentum in the Fulde-Ferrell-Larkin-Ovchinnikov (FFLO) state exhibiting exactly two distinct stable orientations: one orthogonal (FFLO-ADG-O) and the other parallel (FFLO-ADG-P) to the symmetry axis of the pairing gap. The FFLO-ADG-O state dominates at low asymmetries, while the FFLO-ADG-P state prevails at high asymmetries. Additionally, these analysis of normal-superfluid phase separation reveals that the angular dependence of the pairing gap eliminates phase separation in the low-asymmetry regime, whereas the Cooper pair momentum effectively suppresses phase separation at high asymmetries. These two mechanisms act in concert to significantly prevent the occurrence of normal-superfluid phase separation across the entire phase diagram, ensuring the stability of the homogeneous superfluid state over a broad range of asymmetries. These results provide new insights into the interplay between the angular dependence of the pairing gap and the stability of superfluidity in asymmetric nuclear matter.

	\end{abstract}
	\maketitle
	\section{Introduction}

        Neutron-proton ($n-p$) pairing properties are pivotal in various physical contexts. They elucidate the mechanism of deuteron formation in heavy-ion collisions \cite{Baldo1995} and supernovae \cite{Heckel2009, Typel2010, Stein2012} at intermediate energies. Additionally, $n-p$ pairing is integral to understanding the structure of finite nuclei with $N \simeq Z$ \cite{Wang2024, Goodman1999,Ropke2000,Frauendorf2014}, offering insights into the underlying properties of nuclear interactions \cite{shang2023}. Furthermore, these properties may influence the cooling and rotation dynamics in nucleon star models, potentially facilitating pion or kaon condensation in nucleon star model \cite{Brown1994}.

        Different from the pairing between like nucleons, $n-p$ pairing crucially depends on the overlap between the neutron and proton Fermi surfaces. In such two-component superfluid systems, mismatched Fermi surfaces arising from isospin asymmetry can significantly suppress the pairing correlations. There has been considerable interest in both experimental \cite{Zwierlein2006,Partridge2006,shin2006} and theoretical \cite{Sheehy2006,Hu2006,Chen2006,Pao2006,He2006,wang2018,Shang2022} studies of similar asymmetric two-component superfluid systems, such as ultra-cold atomic Fermi gases and neutral dense quark matter, due to the emergence of exotic quantum phenomena. These include nonzero Cooper pair momentum \cite{Fulde1964,Larkin1964}, gapless excitations \cite{Sarma1963}, normal-superfluid phase separation (PS) \cite{shin2006,Sheehy2006,Bedaque2003}, and superfluid density instability \cite{Pao2006,Wu2003,he200612}. An essential origin of these exotic phenomena  is the necessity of accommodating excess unpaired particles in the phase space near the Fermi surface, which significantly influences the system's quantum properties. Unlike $S$-wave pairing, non-$S$-wave pairing features nodes and zero lines in the pairing gap, which provide additional phase space to accommodate excess particles \cite{Wang2015}. This mechanism can mitigate the adverse effect from the asymmetry on pairing correlations. Actually, our previous works \cite{Shang2013} have demonstrated that the angle-dependent gap, resulting from the anisotropic nature of non-$S$-wave pairing, effectively reduces the suppression of the pairing gap caused by the mismatched Fermi surfaces. 
        
        Additionally, two-component $D$-wave Fermi superfluids with density imbalance exhibit a more intricate phase diagram, particularly displaying distinct characteristics in normal-superfluid phase separation \cite{Shang2022}， compared to $S$-wave superfluids \cite{Chen2006}.        
        In particular, the dominant component of the pairing $n-p$ in asymmetric nuclear matter is the attractive isospin-singlet channel $^3SD_1$, which governs the interaction of the pairing $n-p$ below the nuclear saturation density \cite{Alm1993,Sedrakian2000,Lombardo2001,Elgaroy1998}.        
        The pairing gap in this $^3SD_1$ channel represents a hybrid of $S$-wave and $D$-wave pairing, adriven by the tensor force, and exhibits unique features of pairing. Exploring this hybrid structure may provide new insight into the phase diagram.        
        Indeed, prior studies \cite{Shang2015} have demonstrated that the orientation of Cooper pair momentum in Fulde–Ferrell–Larkin–Ovchinnikov (FFLO) state, as originally proposed by Fulde and Ferrell \cite{Fulde1964} and Larkin and Ovchinnikov \cite{Larkin1964}, significantly differs for $^3SD_1$ pairing compared to $S$-wave \cite{He2006,Sedrakian2001} and $D$-wave pairing \cite{Shang2022,Zhang2019}. However, the dynamics of normal-superfluid phase separation for this hybrid structure remains poorly understood. Importantly, the phase separation in this case is expected to differ distinctly from that of both $S$-wave and $D$-wave pairings, highlighting the need for further investigation.

        In this work, we investigate the properties of the $n-p$ pairing in the $^3SD_1$ channel for asymmetric nuclear matter, focusing on the effects of temperature on phase structures. Particular attention is given to the stability regions of the homogeneous superfluid and the FFLO states against PS in the temperature-asymmetry plane. To capture the influence of the hybrid structure of the pairing gap (mixing of $S$-wave and $D$-wave), an angle-dependent gap is adopted. The paper is organized as follows. Section \ref{formalism} outlines the formalism, covering the thermodynamic stability condition of the system, the superfluid density, and the criteria for the stability of homogeneous states against phase separation. Section \ref{results and discussions} presents the results and provides a detailed discussion. Section \ref{summary and outlook} concludes the study with a summary of the findings and the outlook for future research. Throughout this paper, natural units $c = \hbar = k_B = 1$ are used. 
         
	\section{Formalism}\label{formalism}
        We can start with the proton/neutron propagator derived from the solution of the Gorkov equation, which can be expressed in the form: 
        \begin{equation}
		\mathbf{G}^{\sigma, \sigma^{\prime}}_{p/n}\left(\mathbf{k}, \omega_{m}\right)=-\delta^{\sigma, \sigma^{\prime}} \frac{\imath \omega_{m}+\xi_{\mathbf{k}} \mp \delta \varepsilon_{\mathbf{k}}}{\left(\imath \omega_{m}+E_{\mathbf{k}}^{+}\right)\left(\imath \omega_{m}-E_{\mathbf{k}}^{-}\right)}.
	  \end{equation}  
        Additionally, the neutron-proton anomalous propagator is given by:
        \begin{equation}\label{function-propagator}
		\mathbf{F}^{\dag}(\mathbf{k},\omega_{m})=-\frac{\Delta^{\dag}(\mathbf{k})}
		{(i\omega_{m}+E_{\mathbf{k}}^{+})(i\omega_{m}-E_{\mathbf{k}}^{-})}.
	  \end{equation}
        Here, $\omega_{m}$ is the Matsubara frequency and $\Delta(\mathbf{k})$ is the pairing gap matrix defined in the spin space, which possesses the unitary property $\Delta(\mathbf{k})\Delta^{\dag}(\mathbf{k})=ID^{2}(\mathbf{k})$ in the sole $^3SD_1$ channel ($I$ is the identity matrix and $D^2(\mathbf{k})$ is a scalar quantity in spin space). The quasiparticle excitation spectra correspond to the pole of the propagators in the Gorkov equations, $E_{\mathbf{k}}^{\pm}=\sqrt{\xi_{\mathbf{k}}^{2}+D^{2}(\mathbf{k})}\pm\delta\varepsilon_{\mathbf{k}}$ 
        with $\xi_{\mathbf{k}}=\frac{1}{2}(\varepsilon_p+\varepsilon_n)$ and $\delta\varepsilon_{\mathbf{k}}=\frac{1}{2}(\varepsilon_p-\varepsilon_n)$. Where $\varepsilon_{n/p}=\frac{(\mathbf{q}\pm\mathbf{k})^{2}}{2m}-\mu_{n/p}$ is the single-particle energy spectra of neutron/proton and $\mu_{n/p}$ denotes the chemical potential derived self-consistently from the gap equation and the density constraints. The nonzero center-of-mass momentum $2\mathbf{q}$ represents the Cooper pair momentum in the FFLO state. The summation over the Matsubara frequencies provides the density matrix of particles in the condensate, i.e., the pairing probability: 
        \begin{equation}\label{function-Matsubara}
            \nu(\mathbf{k})=\frac{\Delta(\mathbf{k})}{2 \sqrt{\xi_{\mathbf{k}}^{2}+D^{2}(\mathbf{k})}}\left[1-f\left(E_{\mathbf{k}}^{+}\right)-f\left(E_{\mathbf{k}}^{-}\right)\right],
        \end{equation}       
       and the density constraints:
        \begin{align}\label{function-the number density}
         \rho_{p/n} =\sum_{\mathbf{k}}&n_{p/n}(\mathbf{k}) \nonumber\\
         =\sum_{\mathbf{k}}&\Big\{\frac{1}{2}\Big[1+\frac{\xi_{\mathbf{k}}}{\sqrt{\xi_{\mathbf{k}}^{2}+D^{2}(\mathbf{k})}}\Big]f(E_{\mathbf{k}}^{\pm}) \nonumber\\
		&+\frac{1}{2}\Big[1-\frac{\xi_{\mathbf{k}}}{\sqrt{\xi_{\mathbf{k}}^{2}+D^{2}(\mathbf{k})}}\Big]\Big[1-f(E_{\mathbf{k}}^{\mp})\Big]\Big\}.
	  \end{align}
	  Where $f(x)=\big[1+e^{\frac{x}{T}}\big]^{-1}$ is the well-known
        Fermi-Dirac distribution function.    
        
         The gap equation can be naturally derived from the pairing probability through the relation: $\Delta(\mathbf{k})=\sum_{\mathbf{k}^{\prime}}V(\mathbf{k},\mathbf{k}^{\prime})\nu(\mathbf{k}^{\prime})$, where $V(\mathbf{k},\mathbf{k}^{\prime})$ represents the Nucleon-Nucleon (NN) interaction in the relevant channel. Generally, $\Delta(\mathbf{k})$ in the $^3SD_1$ channel includes two sets of independent components corresponding to the $z$ component of the angular momentum, $m_j=0,1$ \cite{Shang2013}. As proposed in Ref. \cite{Shang2013}, we adopt the axial-symmetric gap solution, considering only the $m_j=0$ components.  Consequently, $D^2(\mathbf{k})$ evolves into 
       \begin{align}\label{function-ADG spectrum}
            &D^{2}(\mathbf{k}) = \Delta_{0}^{2}(k) + \sqrt{2} \Delta_{0}(k) \Delta_{0}(k)\left(3 \cos^{2} \theta - 1\right)  \nonumber\\[2mm]
            &\quad \quad \quad \quad + \Delta_{2}^{2}(k)\left(\frac{3 \cos^{2} \theta + 1}{2}\right).
        \end{align}
        $\Delta_{0}$ and $\Delta_{2}$ represent the gaps in the $^3S_1$ and $^3D_1$ channels, respectively, while $(\theta, \phi)$ denotes the orientation of $\mathbf{k}$ in spherical coordinates. The gap equation then takes the form: 
        \begin{align}\label{function-ADG}
		\left(\begin{array}{l}
			\Delta_{0} \\
			\Delta_{2}
		\end{array}\right)(k) & = \frac{-1}{\pi} \int d k^{\prime} k^{\prime 2}\left(\begin{array}{ll}
			V^{00} & V^{02} \\
			V^{20} & V^{22}
		\end{array}\right)\left(k, k^{\prime}\right) \nonumber\\[2mm]
		&\times 
		\int \frac{d \Omega_{\mathbf{k}^{\prime}}}{4\pi} \frac{1-f\left(E_{k^{\prime}}^{+}\right)-f\left(E_{k^{\prime}}^{-}\right)}{\sqrt{\xi_{\mathbf{k}^{\prime}}^{2}+D^{2}\left(\mathbf{k}^{\prime}\right)}}\nonumber\\[2mm]
            &\times\left(\begin{array}{ll}
			f_\theta & g_\theta \\
			g_\theta & h_\theta
		\end{array}\right) 
		\left(\begin{array}{c}
			\Delta_{0} \\
			\Delta_{2}
		\end{array}\right)\left(k^{\prime}\right) .
	\end{align}
      Here, $V^{00}$, $V^{02}$, $V^{20}$, and $V^{22}$ denote the NN interactions in the coupled $^3SD_1$ channel. In particular, the off-diagonal elements $V^{02}$ ($V^{20}$) arise from the tensor force leading to a distinctive mixing of the $S$-wave and $D$-wave pairing gap. The angle matrix, defined as：
      \begin{align}
      \left(\begin{array}{ll}
			f_\theta & g_\theta \\
			g_\theta & h_\theta
		\end{array}\right)=\left(\begin{array}{ll}
			\ \ \ \ \ \ 1 & \ \ \frac{3 \cos^{2} \theta - 1}{\sqrt{2}} \\
			\frac{3 \cos^{2} \theta - 1}{\sqrt{2}} & \ \ \frac{3 \cos^{2} \theta + 1}{2}
		\end{array}\right)
      \end{align} stems from the directional dependence of the anisotropic pairing gap.      
      In early studies of superfluidity in nuclear matter, an angle averaging procedure was often employed to simplify the calculations \cite{Baldo1995,Ropke2000,Lombardo2001,Sedrakian2001}, where $D^{2}(\mathbf{k})$ is replaced by $\frac{1}{4\pi}\int d \Omega_{\mathbf{k}}D^{2}(\mathbf{k})$. Consequently, the angle matrix transforms into the identity matrix. We refer to the gap obtained under this angle-averaging procedure as the angle-averaged gap (AAG), while the gap without angle-averaging is termed the angle-dependent gap (ADG). We should stress here that, for comparison with the results in other references, the pairing gap refers to $\Delta(k)=\sqrt{\Delta_0(k)^2+\Delta_2(k)^2}$.

      It is important to note that the quasiparticle spectrum becomes non-degenerate with respect to the orientation of the Cooper pair momentum when the anisotropic gap is taken into account. To incorporate this in the FFLO state, special attention should be paid to the symmetric and asymmetric parts of the spectrum:  
        \begin{align}\label{function-spectrum}
		\xi_{\mathbf{k}} & =\frac{\mathbf{k}^{2}}{2 m}+\frac{\mathbf{q}^{2}}{2 m}-\mu, \nonumber\\
		\delta \varepsilon & =h-\frac{\mathbf{k} \cdot \mathbf{q}}{m}=h-\frac{k q}{m} \cos (\widehat{\mathbf{k q}}) \nonumber\\
		& =h-\frac{k q}{m}\left[\sin \theta_{0} \sin \theta \cos \left(\phi-\phi_{0}\right)+\cos \theta_{0} \cos \theta\right].
	\end{align}
        Here, $(\theta_0,\phi_0)$ denotes the orientation of $\mathbf{q}$ in spherical coordinates. We should note that the properties of the spherical harmonics have been adopted to expand the cosine of the angle between $2\mathbf{q}$ and $\mathbf{k}$. The term $\frac{\mathbf{k} \cdot \mathbf{q}}{m}$ associated with the Cooper pair momenta can mitigate the suppression caused by mismatched Fermi surfaces $h$ in certain directions. We emphasize that both the densities \ref{function-the number density} and the gap equation \ref{function-ADG} are independent of $\phi_{0}$. Specifically, $\phi_{0}$ can be eliminated by selecting a special spherical coordinate system where the direction of $\mathbf{q}$ is expressed as $(\theta_0,\phi_0=0)$. In this scenario, only $\theta_0$ serves as a parameter to describe the direction of the Cooper pair momentum. Consequently, both the magnitude $q$ and the orientation $\theta_0$ of the Cooper pair momentum must be determined by minimizing the system's energy. 
        
        For asymmetric nuclear matter with total number density $\rho=\rho_n + \rho_p$ and isospin asymmetry $\alpha=(\rho_n - \rho_p)/\rho$ at finite temperature $T$, the essential thermodynamic quantity describing the system is the free energy defined as: 
        \begin{equation}\label{function-free energy}
		\mathcal{F} = \Omega  + \mu\rho + h \delta\rho,
	  \end{equation}
        where $\delta\rho=\rho_n - \rho_p$ and the thermodynamic potential $\omega$ takes the form:
        \begin{equation}
		\Omega = U  - TS.
	  \end{equation}
        The entropy of the superfluid state can be expressed as:
        \begin{align}\label{function-entropy}
		S =& -2\sum_{\mathbf{k}}\Big\{f(E_{\mathbf{k}}^+)\ln f(E_{\mathbf{k}}^+) + \big[1-f(E_{\mathbf{k}}^+)\big]\ln\big[1-f(E_{\mathbf{k}}^+)\big] \nonumber\\[2mm]
		&+ f(E_{\mathbf{k}}^-)\ln (E_{\mathbf{k}}^-) + \big[1-f(E_{\mathbf{k}}^-)\big]\ln\big[1-f(E_{\mathbf{k}}^-)\big]\Big\},
	  \end{align}
        while the internal energy of the grand canonical ensemble is given by:
        \begin{align}\label{function-interal energy}
		U = \sum_{\mathbf{k}}\Big[\varepsilon_n n_n(\mathbf{k}) +\varepsilon_p n_p(\mathbf{k})\Big] - \sum_{\mathbf{k},\mathbf{k}^{\prime}}V(\mathbf{k},\mathbf{k}^{\prime})\nu^{\dag}(\mathbf{k})\nu(\mathbf{k}^{\prime}).
	   \end{align} 
        The second term of the internal energy accounts for the BCS mean-field interaction between particles in the condensate. By utilizing the pairing probability \ref{function-Matsubara}, the density constraints \ref{function-the number density}, and the gap equation \ref{function-ADG}, the thermodynamic potential can be expressed as:
	\begin{align}\label{function-thermodynamic potential}
		\Omega =& \sum_{\mathbf{k}}\frac{D(\mathbf{k})^2}{\sqrt{\xi_{\mathbf{k}}^2+D(\mathbf{k})^2}}\big[1-f(E_{\mathbf{k}}^+)-f(E_{\mathbf{k}}^-)\big]\nonumber \\[2mm]
		& - 2\sum_{\mathbf{k}}\Big[ \sqrt{\xi_{\mathbf{k}}^{2}+D^{2}(\mathbf{k})}-\xi_{\mathbf{k}} + T\ln(1+e^{-E_{\mathbf{k}}^-/T})\nonumber \\[2mm] &+T\ln(1+e^{-E_{\mathbf{k}}^+/T})\Big].
	  \end{align}
         The thermodynamically stable solution should minimize the difference between the free energies of the superconducting and normal states, defined as $\delta\mathcal{F}=\mathcal{F}_{S}-\mathcal{F}_{N}$, where the free energy in the normal state follows from Eq. \ref{function-free energy} as $\Delta\rightarrow 0$. Thus, the magnitude $q$ and the orientation $\theta_0$ of the Cooper pair momentum can be determined by the conditions:
        \begin{equation}\label{function-q}
		\frac{\mathcal{D}\mathcal{F}}{\mathcal{D}q} = 0, \ \ \frac{\mathcal{D}\mathcal{F}}{\mathcal{D}\theta_{0}} = 0,
	  \end{equation}
        with the definitions:
        \begin{align}
		 \frac{\mathcal{D}}{\mathcal{D}q}&=\frac{\partial}{\partial q}+\frac{\partial\mu}{\partial q}\frac{\partial}{\partial \mu}+\frac{\partial h}{\partial q}\frac{\partial}{\partial h}, \nonumber \\
        \frac{\mathcal{D}}{\mathcal{D}\theta_{0}}&=\frac{\partial}{\partial \theta_{0}}+\frac{\partial\mu}{\partial \theta_{0}}\frac{\partial}{\partial \mu}+\frac{\partial h}{\partial \theta_{0}}\frac{\partial}{\partial h}.
	  \end{align}
        Using the relations , it follows that $\mathcal{D}\mathcal{F}/\mathcal{D}q=\partial\Omega/\partial q$ and $\mathcal{D}\mathcal{F}/\mathcal{D}\theta_{0}=\partial\Omega/\partial \theta_{0}$. It is important to note that Eq. \ref{function-q} should be resolved self-consistently with the density constraints \ref{function-the number density} and the gap equation \ref{function-ADG} to determine the value $q$ and $\theta_{0}$.

	\subsection{Superfluid density}
         The superfluid density, which quantifies the instability induced by a tiny velocity $\mathbf{v}_{s}$, must be positive for the system to remain a stable state \cite{Pao2006,Pao200674}. By employing linear response theory, the superfluid density can be calculated by analyzing the system's response to an fictitious external vector potential. For canonical ensemble, the supercurrent $\mathbf{j}_{s}$ and the superfluid density $\rho_{s}$ can be defined via the Taylor expansion of free energy $\mathcal{F}$ near $\mathbf{q}_{s}=(q_{s},\theta_{0s}, \phi_{0s})$ [here supposing $(q_{s},\theta_{0s})$ minimize the free energy \ref{function-free energy} and satisfy the density constraints \ref{function-the number density} and gap equation \ref{function-ADG}],
         \begin{equation}
		\mathcal{F}(\mathbf{v}_{s}) = \mathcal{F}(0)+\mathbf{j}_{s}\cdot\mathbf{v}_{s}+\frac{1}{2}\rho_{s}\mathbf{v}^{2}_{s}+...
	  \end{equation}
        Where 
        \begin{equation}
		\mathcal{F}(0) = \left.\mathcal{F}\right|_{\mathbf{q}_{s}} , \ \ \left.\mathbf{j}_{s}=m\frac{\mathcal{D}\mathcal{F}}{\mathcal{D}\mathbf{q}}\right|_{\mathbf{q}_{s}}, \ \
        \left.\rho_{s}=m^2\frac{\mathcal{D}^2\mathcal{F}}{\mathcal{D}\mathbf{q}^2}\right|_{\mathbf{q}_{s}}.
	  \end{equation}
        The term $\mathbf{j}_{s}$ vanishes due to the equation for the pair momentum \ref{function-q}. The positive superfluid density is related to the positively definite matrix $\frac{\mathcal{D}^2\mathcal{F}}{\mathcal{D}\mathbf{q}^2}$, which can be expressed in the spherical coordinates as: 
        \begin{align}
		\frac{\mathcal{D}^2\mathcal{F}}{\mathcal{D}\mathbf{q}^2}=\left(\begin{array}{ccc}
		\frac{\mathcal{D}^2\mathcal{F}}{\mathcal{D}q^2} & \frac{\mathcal{D}}{\mathcal{D}q}(\frac{1}{q}\frac{\mathcal{D}\mathcal{F}}{\mathcal{D}\theta_{0}}) & 0 \\ \\
		\frac{\mathcal{D}}{\mathcal{D}q}(\frac{1}{q}\frac{\mathcal{D}\mathcal{F}}{\mathcal{D}\theta_{0}}) & \ \  \frac{1}{q}\frac{\mathcal{D}\mathcal{F}}{\mathcal{D}q}+\frac{1}{q^2}\frac{\mathcal{D}^2\mathcal{F}}{\mathcal{D}\theta^2_{0}} & 0 \\ \\
		0 & 0 &\frac{1}{q}\frac{\mathcal{D}\mathcal{F}}{\mathcal{D}q}+\frac{\cos\theta_{0}}{q^2\sin\theta_{0}}\frac{\mathcal{D}\mathcal{F}}{\mathcal{D}\theta_{0}}
		\end{array}\right).
	\end{align}        
        Since the free energy is independent of the azimuth angle $\phi_{0}$, all derivatives with respect to $\phi_{0}$ vanish. This indicates that the superfuid density matrix, including the $\mathbf{\hat{e}}_{\phi_{0}}\mathbf{\hat{e}}_{\phi_{0}}$ component, is also independent of $\phi_{0}$. For the $S$-wave pairing gap or for the AAG state, $\frac{\mathcal{D}^2\mathcal{F}}{\mathcal{D}\mathbf{q}^2}$ is isotropic. In this case, only the term $\frac{\mathcal{D}^2\mathcal{F}}{\mathcal{D}q^2}$ describes the susceptibility related to the emergence of the Cooper pair momentum, commonly refers to the pair momentum susceptibility \cite{He2006,Giannakis2005}. However, for the ADG state,
        the anisotropic pairing gap gives rise to an additional significant term $\frac{\mathcal{D}^2\mathcal{F}}{\mathcal{D}\theta_{0}^2}$ which characterizes the orientation instability of the Cooper pair momentum. We will refer to this term as pair momentum orientation susceptibility hereafter.

        In the case $q=0$ without the FFLO state configuration, $\left.\frac{\mathcal{D}^2\mathcal{F}}{\mathcal{D}\mathbf{q}^2}\right|_{\mathbf{q}=0}=\left.\frac{\partial^2\Omega}{\partial\mathbf{q}^2}\right|_{\mathbf{q}=0}$. Thus, the superfluid density develops into:
		\begin{align}\label{function-density tensor}
		\rho_{\mathrm{s}}=\left(\begin{array}{ccc}
		\rho_{\mathrm{T}} \sin ^{2} \theta_{0}+\rho_{\mathrm{L}} \cos ^{2} \theta_{0} & \left(\rho_{\mathrm{T}}-\rho_{\mathrm{L}}\right) \cos \theta_{0} \sin \theta_{0} & 0 \\
		\left(\rho_{\mathrm{T}}-\rho_{\mathrm{L}}\right) \cos \theta_{0} \sin \theta_{0} & \rho_{\mathrm{T}} \cos ^{2} \theta_{0}+\rho_{\mathrm{L}} \sin ^{2} \theta_{0} & 0 \\
		0 & 0 & \rho_{\mathrm{T}}
		\end{array}\right),
	\end{align}
	 in spherical coordinates. Due to the axial symmetry of the pairing gap adopted in the current paper, the anisotropic superfluid density tensor can be decomposed into transverse and longitudinal components, resulting from the $O(2)$ symmetry. The transverse and the longitudinal parts are defined as follow:
	\begin{equation}\label{function-density L&T}
		\rho_L = m\rho + \int \frac{k^4}{4\pi^2} \int_{-1}^{1} \cos^2 \theta  \mathrm{~d}\cos\theta (f'(\tilde{E}_{\mathbf{k}}^+) + f'(\tilde{E}_{\mathbf{k}}^-)) \nonumber,
	\end{equation}
	\begin{equation}
		\rho_T = m\rho + \int \frac{k^4}{8\pi^2} \int_{-1}^{1} \sin^2 \theta  \mathrm{~d}\cos\theta (f'(\tilde{E}_{\mathbf{k}}^+) + f'(\tilde{E}_{\mathbf{k}}^-)).
	\end{equation}
        where $\tilde{E}_{\mathbf{k}}^{\pm}=\left.E_{\mathbf{k}}^{\pm} \right|_{q=0}$. When adopting the AAG state or equivalently considering the pairing gap as the $S$-wave pairing, we find that $\rho_s=\rho_L = \rho_T$, indicating an isotropic superfluid density. In contrast, in the ADG scenario, the discrepancy between $\rho_L$ and $\rho_T$ results in an anisotropic superfluid density tensor $\rho_{s}$. Moreover, a negative $\rho_s$ implies instability in the conventional BCS state and the potential emergence of an FFLO state. 
        
        \subsection{Stability condition against phase separation in the homogeneous superfluid state}
        
        In addition to the Cooper pair momentum, an inhomogeneous mixed phase composed of normal and superfluid components could also mitigate the suppression caused by asymmetry \cite{Bedaque2003,Viverit2000,Cohen2005}. To evaluate the stability of the homogeneous superfluid state in asymmetric nuclear matter, one can investigate the response of total free energy $F=\int d^3\mathbf{r}\mathcal{F}[\rho_\tau(\mathbf{r})]$(where $\tau=n,p$) to an density distribution fluctuation $\delta\rho_{\tau}(\mathbf{r})$ that deviate from a homogeneous distribution. Consequently, the first-order variation $\delta F$ vanishes, leaving the second-order variation $\delta^2F$ takes the quadratic form,
	\begin{equation}\label{function−variation}
		\delta^{2} F=\frac{1}{2} \int \mathrm{d}^{3} \mathbf{r} \sum_{\tau, \tau'=n, p} \frac{\partial^{2} \mathcal{F}}{\partial \rho_{\tau} \partial \rho_{\tau'}} \delta \rho_{\tau}(\mathbf{r}) \delta \rho_{\tau'}(\mathbf{r}).
	\end{equation}
        To ensure a stable homogeneous sole phase, the $2 \times 2$ matrix 
        $\partial^{2} \mathcal{F} / \partial \rho_{\tau} \partial \rho_{\tau'}=\partial\mu_\tau' / \partial\rho_\tau$ must be positive definite. Note that  
        $\partial / \partial\rho_a=\partial / \partial\rho+\partial / \partial\delta\rho$, 
        $\partial / \partial\rho_b=\partial / \partial\rho-\partial / \partial\delta\rho$,
        the positive definiteness of the matrix $\partial^{2} \mathcal{F} / \partial \rho_{\tau} \partial \rho_{\tau'}$ is equivalent to the positive definiteness of the matrix $\mathcal{M}$ defined as:	
	   \begin{align}\label{function-M22}
		\mathcal{M}=\left(\begin{array}{cc}
			\frac{\partial\mu}{\partial\rho} & \frac{\partial\mu}{\partial\delta\rho} \\
			\frac{\partial h}{\partial\delta\rho} & \frac{\partial h}{\partial\delta\rho}
		\end{array}\right).
	  \end{align}
        This implies that both eigenvalues $\lambda_{\pm}$ of $\mathcal{M}$ are positive, given by, 
	  \begin{equation}\label{function−eigenvalue}
		\lambda_{ \pm}=\frac{\operatorname{Tr}\mathcal{M} \pm \sqrt{\operatorname{Tr}\mathcal{M}^{2}-4 \operatorname{det}\left(\mathcal{M}\right)}}{2}.
	  \end{equation}
        Here, the trace $\operatorname{Tr}\mathcal{M}=\frac{\partial\mu}{\partial\rho}+\frac{\partial h}{\partial\delta\rho}$. Fortunately, $\lambda_+$, which corresponds to the stability of the total density distribution, is always positive. The other eigenvalue $\lambda_-$ is roughly associated with the stability of the isospin asymmetry distribution and ultimately determines the stability condition against phase separation in the homogeneous superfluid state.
	
	\section{Results and Discussions}\label{results and discussions}
           To focus on the phase structure of neutron-proton superfluid for asymmetric nuclear matter in the temperature-asymmetry plane, the bare $NN$ interaction, specifically the Argonne $v18$ potential, is adopted as the pairing interaction for simplicity. This approach neglects the screening effects on the pairing interaction, which remain an open problem for neutron-proton pairing despite several studies \cite{Schulze1996,shen2005,cao2006,zhang2016}. Furthermore, the free single-particle (s.p.) spectrum is utilized to facilitate the differential calculation of the free energy. While incorporating self-energy effects would provide more realistic results, it poses significant computational challenges \cite{Fan2019} to obtain the second differential term of free energy. However, although these two approximations may affect the magnitude of the pairing gap, the overall structure of the phase diagram in the temperature-asymmetry plane is unlikely significantly changed. In the present calculations, the net density is fixed at the empirical saturation density of nuclear matter, $\rho_{0}=0.17fm^{-3}$.
           
	\subsection{Phase diagram with  $\mathbf{q}=0$ }        
\begin{figure}[H]
			\centering
			\includegraphics[width=0.7\linewidth]{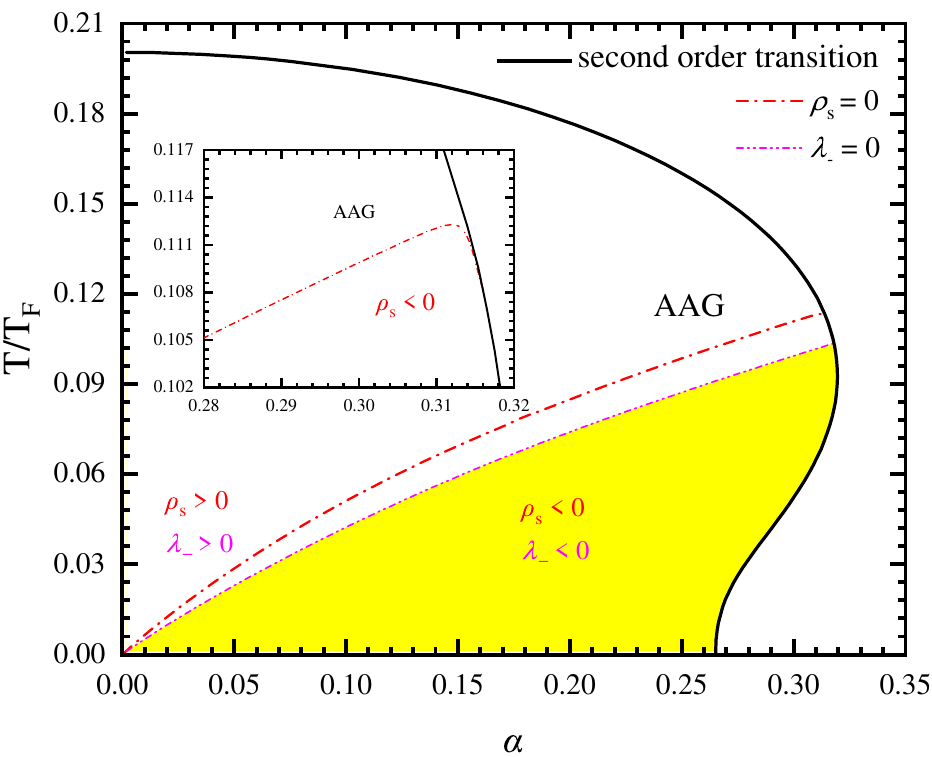}
			\caption{Phase diagram of the AAG state in the $T-\alpha$ plane. The solid line represents the second-order phase transition from the superfluid to the normal states. The yellow-shaded region corresponds to the PS and the magenta dash-dot-dot line indicates the stability boundary against PS. The dash-dotted line marks the zero value of the superfluid density. The inset highlights the details of the superfluid density near the temperature range $0.1092 < \frac{T}{T_F} < 0.114$.}
			\label{fig-aag}
		\end{figure}

\begin{figure}[H]
			\centering
			\includegraphics[width=0.7\linewidth]{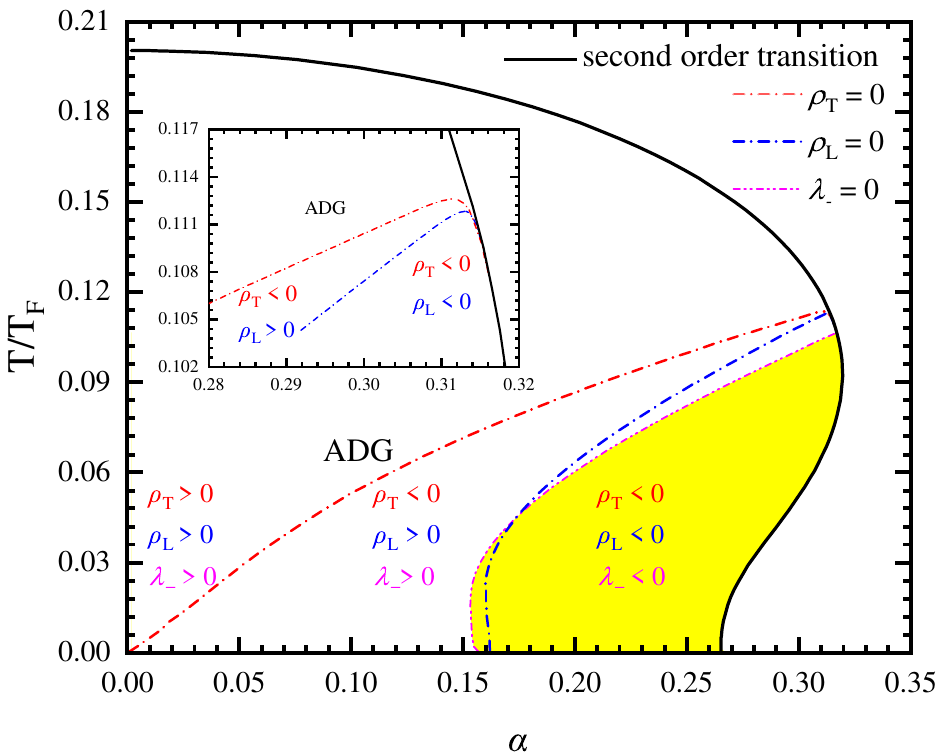}
			\caption{Phase diagram of the ADG state in the $T-\alpha$ plane. Similar to Fig. \ref{fig-aag}，expect that the superfluid density decomposes into the transverse ($\rho_{T}$) and longitudinal ($\rho_{L}$) components, represented by the red and blue dash-dotted lines, respectively.}
			\label{fig-adg}
		\end{figure}

        We first analyze the scenario without pair momentum. The phase diagrams for the AAG and ADG states are exhibitted in Figures \ref{fig-aag} and \ref{fig-adg}, respectively. Hereafter, the temperature in all the figures is normalized by the Fermi temperature, defined as $T_F=\frac{k^2_F}{2m}$ with the average Feimi  momentum $k_F=(3\pi^2\rho/2)^{\frac{1}{3}}$.

            Following the angle-averaging procedure, the pairing gap becomes equivalent to the $S$-wave pairing. Consistent with the case of $S$-wave pairing at fixed number densities \cite{He2006,He2006214527}, the phase transition from the AAG state to the normal state is identified as a second-order phase transition, as illustrated in Fig. \ref{fig-aag}. This behavior is further corroborated by the fact that the pairing gap vanishes as the system approaches the critical asymmetry, $\alpha_{c}$. Interestingly, $\alpha_{c}$ initially increases and then decreases with rising temperature. This trend arises because neutron-proton (n-p) pairing depends crucially on the overlap between neutron and proton Fermi surfaces. At zero temperature, pairing is suppressed due to the mismatched Fermi surfaces. As the temperature increases, thermal excitations enhance the occupancy probabilities of neutrons and protons, resulting in a larger critical asymmetry. However, with further temperature increases, the superfluid state is eventually disrupted by thermal dispersion. 
            
            The superfluid density $\rho_s$, represented by the dash-dotted line in Fig. \ref{fig-aag},  measures the tendency for the emergence of the Cooper pair momentum. Notably, a peak of the $\rho_s=0$ line appears near the temperature range $0.109 < \frac{T}{T_F} < 0.114$, indicating the emergence and subsequent disappearance of Cooper pair momentum with increasing asymmetries in this temperature range, as detailed in Section \ref{section-fflo}. This phenomenon may arise from the tensor component of the pairing interaction. Additionally, the stability condition against PS in the AAG state is depicted by the dash-dot-dot line, while the yellow-shaded region corresponds to the PS. At zero temperature, the entire asymmetry range falls within the normal-superfluid phase separation region.
    
            Once the angle-dependent gap is adopted, as shown in Fig. \ref{fig-adg} the superfluid density decomposes into the transverse ($\rho_{T}$) and longitudinal ($\rho_{L}$) components, corresponding to the tendency for the appearance of Cooper pair momentum in the directions orthogonal and parallel to the symmetry axis of ADG, respectively. Consistent with the discussion in Ref. \cite{Shang2013,Shang2022}, $\rho_{T}$ reaches a negative value first with increasing asymmetry, signaling the initial emergence of the FFLO-ADG-Orthogonal state.
            
            Compared with the results in the AAG state, the normal-superfluid PS region is significantly suppressed, particularly at low temperatures. As is well known, asymmetry necessitates that excess unpaired particles occupy certain phase-space region near the Fermi surface, which is severely detrimental to S-wave pairing. In contrast, for non-$S$-wave pairing, gapless nodes and zero lines exist in the phase space near the Fermi surface, providing a mechanism to accommodate these excess particles. This facilitates pairing near the average Fermi surface and significantly mitigates the instabilities inherent to $S$-wave pairing with number density imbalance, including the transition from homogeneous superfluid phases to inhomogeneous phases. As the asymmetry increases, this mechanism becomes insufficient to counteract the suppression effect of the asymmetry on pairing and the critical asymmetry $\alpha_{c}$ shows a trend consistent with the results for the AAG state. Moreover, with increasing temperature, thermal excitations further weaken the mitigating effect of the ADG mechanism, leading to an expanded relative asymmetry range where normal-superfluid PS occurs.

          \begin{figure}[H]
			\centering
			\includegraphics[width=1.0\linewidth]{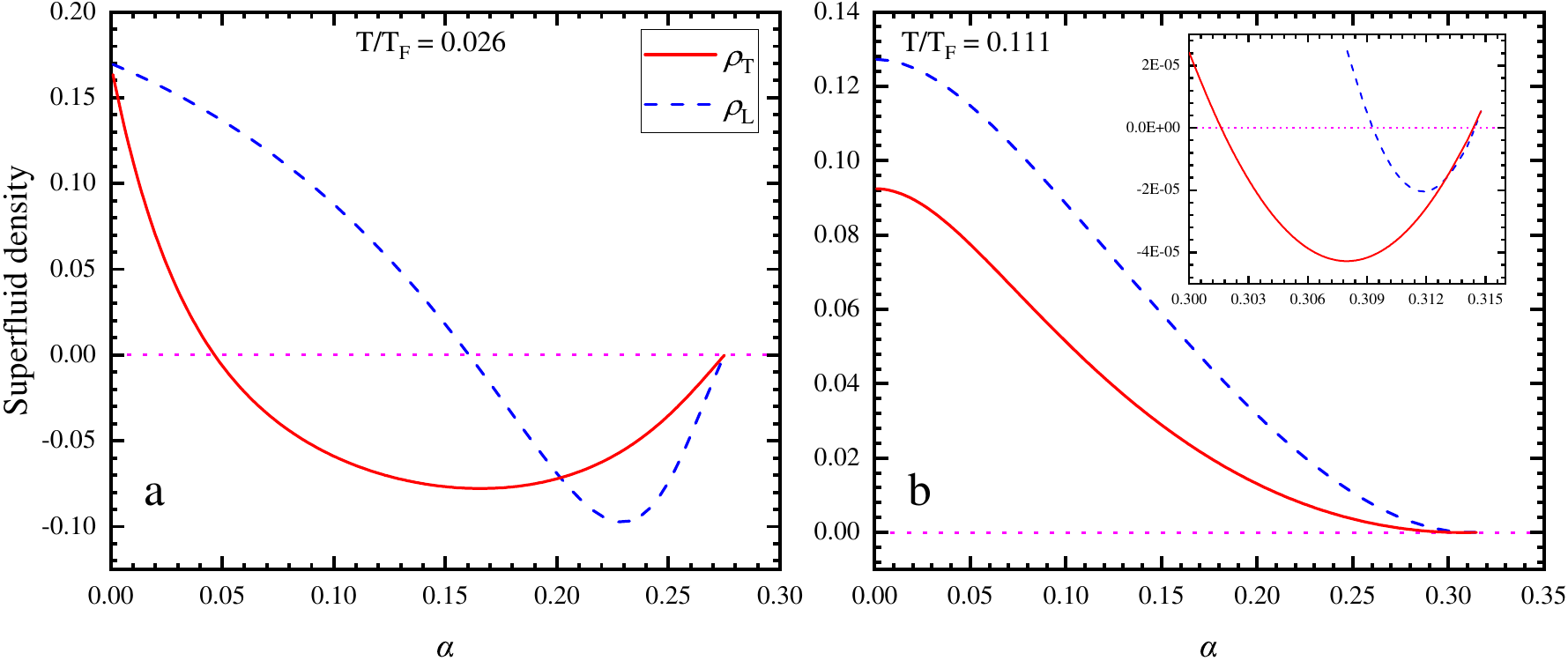}
			\caption{The transverse ($\rho_{T}$) and longitudinal ($\rho_{L}$) components of superfluid density as a function of the isospin asymmetry, represented by the red solid and blue dashed lines, respectively. The left panel corresponds to the temperature $\frac{T}{T_F} = 0.026$, while the right panel corresponds to $\frac{T}{T_F} = 0.111$. The inset in the right panel provides details of the superfluid density near the critical asymmetry.}
			\label{fig-density adg}
		\end{figure}
                Notably, the peak of the $\rho_{T}/\rho_{L}=0$ line near the temperature range $0.109 < \frac{T}{T_F} < 0.114$ is also observed in Fig. \ref{fig-adg}. To further elucidate this phenomenon, the transverse ($\rho_{T}$) and longitudinal ($\rho_{L}$) components of the superfluid density at two distinct temperatures are presented in Fig. \ref{fig-density adg}. At shown in the left panel, both $\rho_{T}$ and $\rho_{L}$ cross zero line once and vanish at the critical asymmetry $\alpha_{c}$. In contrast, in the right panel, these two components cross the zero line twice, corresponding to the appearance of the observed peak.
		
	\subsection{Phase diagram of FFLO state}\label{section-fflo}
\begin{figure}[H]
			\centering
			\includegraphics[width=0.7\linewidth]{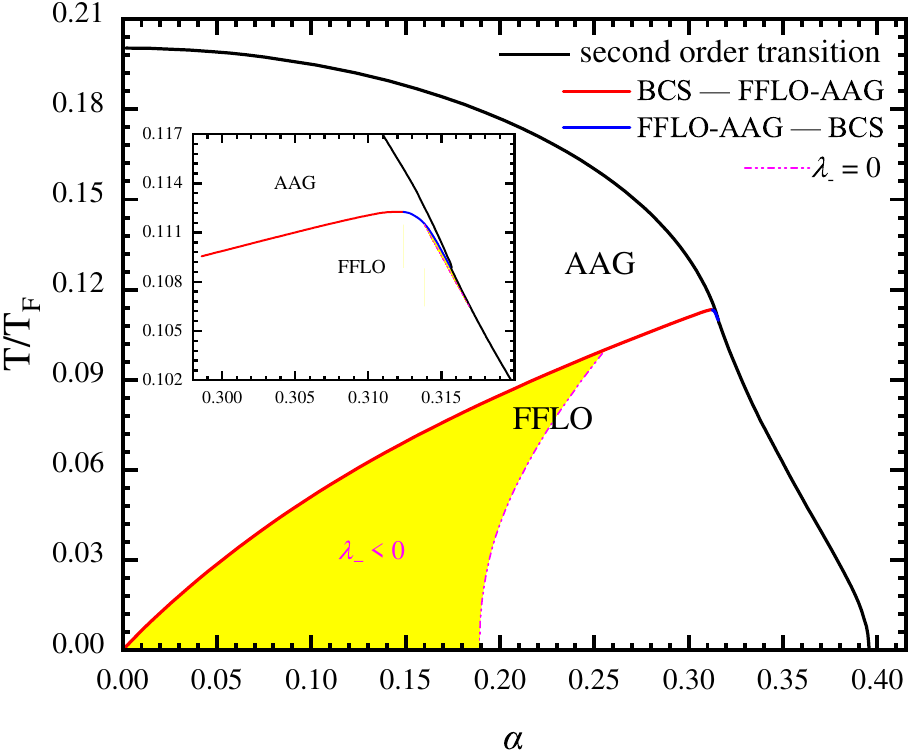}
			\caption{Phase diagram of the FFLO-AAG state in the $T-\alpha$. The black, red and blue solid line represent the second-order phase transition from the superfluid state to the normal state, the conventional BCS state to the FFLO-AAG state and the FFLO-AAG state back to the conventional BCS state, respectively. The yellow-shaded regions indicate the presence of PS, while the magenta dash-dot-dot lines denote the stability boundary against PS.  The inset offers a  detailed view of the special phase structure near the temperature range $0.109 < \frac{T}{T_F} < 0.114$.}
			\label{fig-fflo aag}
		\end{figure}

\begin{figure}[H]
			\centering
			\includegraphics[width=0.7\linewidth]{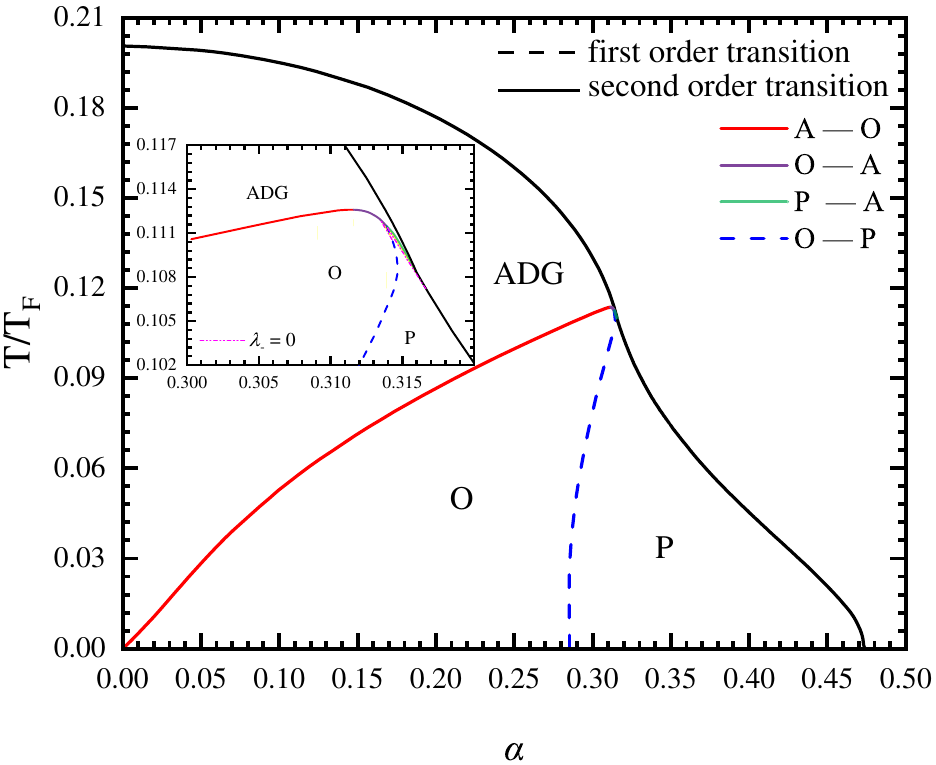}
			\caption{Phase diagram of the FFLO-ADG state in the $T-\alpha$. The dashed lines represent the first-order phase transition, while the solid lines indicate the second-order phase transition. The yellow-shaded region denotes the the presence of PS, while the magenta dash-dot-dot line marks the stability boundary against PS. The inset details the intricate phase structure near the the temperature range $0.109 < \frac{T}{T_F} < 0.114$. The red solid, purple solid, green solid, and blue dashed lines represent the phase transition from the ADG state to the FFLO-ADG-O state (A-O), from the FFLO-ADG-O state back to the ADG state (O-A), the FFLO-ADG-P state back to ADG state (P-A), and the FFLO-ADG-O state to the FFLO-ADG-P state (O-P), respectively.}
			\label{fig-fflo adg}
		\end{figure}

           Carefully accounting for the angular dependence of the pairing gap can effectively eliminate the emergence of the normal-superfluid PS at low asymmetries. While the introduction of Cooper pair momentum prevents the occurrence of normal-superfluid PS at higher asymmetries \cite{He2006,Chen2006}. Figures \ref{fig-fflo aag} and \ref{fig-fflo adg} illustrate the phase diagrams for the FFLO state with angle-averaged gap and angle-dependent gap, referred to hereafter as FFLO-AAG and FFLO-ADG, respectively.

            The phase diagram of FFLO-AAG state, shown in Fig. \ref{fig-fflo aag}, closely resembles that of $S$-wave pairing \cite{He2006}, as the angle-averaged gap is equivalent to the $S$-wave pairing. The FFLO state is identified exactly right at $\rho_{s}=0$ in the AAG state, indicating a second-order phase transition from the conventional BCS state to the FFLO-AAG state. The emergence of Cooper pair momentum in the FFLO-AAG state effectively prevents normal-superfluid phase separation at higher asymmetries. Additionally, the presence of Cooper pair momentum significantly expands the superfluid region in the phase diagram. It is important to highlight that near the temperature range $0.109 < \frac{T}{T_F} < 0.114$, a sequence of second-order transitions occurs: first, from the conventional BCS state to the FFLO-AAG state, and subsequently back to the conventional BCS state. This distinctive phase structure arises from the anomalous behavior of $\rho_{s}=0$, as illustrated in the inset of Fig. \ref{fig-aag}. Furthermore, near the transition from the FFLO-AAG state back to the conventional BCS state, a tiny, narrow region normal-superfluid PS region is observed, which merits particular attention.
           
            Finally, we turn to the phase diagram of the FFLO-ADG state, as depicted in Fig. \ref{fig-fflo adg}. Due to the anisotropic nature of the pairing gap, the FFLO state is nondegenerate with the orientation of Cooper pair momentum. As discussed in Refs. \cite{Shang2015,Zhang2019,Shang2022}, two specific directions, corresponding to $\theta_{0}=0$ and $\theta_{0}=\pi/2$, are favored for the ground state. These two configurations are hereafter referred to as the FFLO-ADG-P state and the FFLO-ADG-O state, respectively.
            Beyond these two orientations, we systematically calculated the free energy for other directions and found that only these two orientations are thermodynamically stable. Moreover, the calculation of the pair momentum orientation susceptibility $\left.\frac{\mathcal{D}^2\mathcal{F}}{\mathcal{D}\theta_{0}^2}\right|_{\theta_{0}=0,\pi/2}=\frac{\partial^2\mathcal{F}}{\partial\theta_{0}^2}$, consistently yields positive values for the obtained solution, further confirming the thermodynamic stability of these two orientations.
            
            Consistent with the superfluid density analysis in the ADG state, the phase transition from the superfluid state to the normal state evolves into a sequence of transitions: initially, a second-order transition from the ADG state to the FFLO-ADG-O state; subsequently, a first-order transition from the FFLO-ADG-O state to the FFLO-ADG-P state; and finally, a second-order transition from the FFLO-ADG-P state to the normal state. In comparison to the phase diagram of the FFLO-AAG state, the angular dependence of the pairing gap significantly extends the asymmetry range over which the FFLO state exists. Moreover, it is remarkably evident that the normal-superfluid PS region almost completely disappears. The configurations of the ADG state and the FFLO state eliminate the normal-superfluid PS region at relatively low and high asymmetries, respectively. When these features combine in the FFLO-ADG state, the homogeneous superfluid state appears to remain stable against phase separation throughout the entire existing region. As in the FFLO-AAG state, the phase diagram becomes more intricate near the temperature range $0.109 < \frac{T}{T_F} < 0.114$. In addition to the aforementioned transitions, second-order transitions from the FFLO-ADG-O state and the FFLO-ADG-P state back to the ADG state are observed, as detailed in the inset. Furthermore, a tiny, narrow region normal-superfluid PS region is also identified.   

	\section{Summary and outlook}\label{summary and outlook}
        Summarily, we have investigated the phase structure of the neutron-proton superfluid in asymmetric nuclear matter, focusing particularly on normal-superfluid PS that arise from the number density imbalance. The angular dependence of the pairing gap, originating from the $^3SD_1$ channel $NN$ interaction, results in the freedom of the Cooper pair momentum orientation in the FFLO state. This can significantly enrich the phase structure. Briefly, the FFLO state splits into two branches, i.e., the FFLO-ADG-O and FFLO-ADG-P states, located in the relatively low and high asymmetries. The phase transition from the ADG state to the FFLO state and the superfluid state to the normal state are of the second order, while the transition from the FFLO-ADG-O state to FFLO-ADG-P state is a first-order transition. Additionally, near the temperature range $0.109 < \frac{T}{T_F} < 0.114$ an intricate structure appears: an initial transition from the conventional BCS state to the FFLO state and a subsequent transition from the FFLO state back to the conventional BCS state. This special structure is consistent with the behavior of the superfluid density in the conventional BCS state.   

        Most interestingly, our findings reveal that the Cooper pair momentum in the FFLO state can effectively remove the instability against the homogeneous superfluid phase at high asymmetries, while the angular dependence of the pairing gap can eliminate the instability of homogeneous superfluid phase at low asymmetries. Together, these two mechanisms nearly completely prevent the occurrence of normal-superfluid phase separation across the entire phase diagram. Although both the Cooper pair momentum and the angular dependence of the pairing gap mitigate the suppression from asymmetries on two-component superfluid system, the current investigation of the phase diagram allows us to clearly distinguish their respective regions of contribution. This further highlights that the two mechanisms work in cooperation and mutually reinforce each other. 
        
        In this paper, we reveal the impact of the angle-dependent gap on neutron-proton superfluidity in asymmetric nuclear matter within the weak BCS regime, especial focus on the effect of angle-dependent gap on the normal-superfluid PS. As the density decreases, neutron-proton pairing transitions smoothly into the strong BCS regime and may even lead to Bose-Einstein condensation (BEC), resulting in the formation of deuterons. However, the role of the angular dependence of the pairing gap during this transition remains unclear. Moreover, our calculations are based solely on the free energy spectrum, and the contributions from the medium have yet to be incorporated. These aspects will be addressed in future studies.

	\section*{Acknowledgments}
          One of the authors, X. L. Shang, would like to express sincere gratitude for the financial support provided by the Peng Huanwu Center for Theoretical Physics Innovation. This work was supported by the Youth Innovation Promotion Association of Chinese Academy of Sciences (Grant No. Y2021414); the National Natural Science Foundation of China under Grant No. 12375117, and No. 12047503; the Strategic Priority Research Program of Chinese Academy of Sciences under Grant No. XDB34000000; CAS Project for Young Scientists in Basic Research YSBR-088；the Gansu Natural Science Foundation under Grant No. 23JRRA675, and No. 20JR5RA481.
	
	\newpage
	\bibliography{intro,text}
\end{document}